# EXTRAGALACTIC DATABASES: PRESENT AND FUTURE

## H. ANDERNACH

Observatoire de Lyon, F-69651 Saint-Genis Laval Cedex, France

**Abstract** − Current extragalactic databases are reviewed, including object-oriented databases, astronomical catalogues and compilations, as well as image archives and object catalogues from large-scale surveys. One challenge of the future will be to maintain the accessibility of all published information on individual objects; the other will be the cross-identification between various ongoing and planned large-scale surveys, some of them resulting in several hundred million objects each. Powerful user interfaces are being developed to tackle these tasks.

## 1. INTRODUCTION

The term "database" is used in astronomy for almost anything ranging from a file of record cards up to contrieved software systems which translate requests from remote users into a query of an information archive of which the user does not need to know its location. I shall concentrate here on the *content* of *electronically accessible* databases rather than on their *methods of access*. The access is usually provided by *user interfaces*, not to be confused with the actual information stored in the database. I shall divide databases − somewhat arbitrarily − into three groups, (a) the highly-organised and object-oriented databases; (b) published catalogues and compilations (searchable or not); (c) very bulky databases like images of large parts of the sky at any wavelength and object catalogues extracted from them. I shall not discuss the manifold of database systems which allow to browse different selections of astronomical catalogues (often with large overlap between the systems). Neither shall I describe bibliographical databases like the abstract services of ADS or ESIS, or the access to raw or calibrated data from archives of large space- or ground-based astronomical observatories. For more detail on such facilities and their methods of access I refer to a recent review [2].

The term "extragalactic" will be used here in a wider sense of "*potentially* extragalactic", so as to include databases with either mixed types of objects and of objects with yet undetermined nature. Examples of databases of "mixed" objects are SIMBAD at the Strasbourg Astronomical Data Centre (CDS), or the plate scan catalogues drawn from the digitised optical sky surveys. By definition, LEDA (at Lyon and Meudon) [14] and NED at IPAC [6] are databases of only extragalactic objects. However, NED includes a number of unidentified (e.g. radio- or X-ray) sources and also retains (and marks as such) those objects which have been reclassified as Galactic objects after their folding into NED.

## 2. OBJECT-ORIENTED DATABASES

The common feature of such database systems is the continuous attempt of their managers to attach newly published information on a given object to an entry already existing in the database, or - if it does not exist - to create a new entry. As such, they provide an efficient tool to find published information on an object, to be queried either by (one of its) names or by its position in the sky, in any of the commonly used coordinate systems.

Table 1 summarizes and compares the contents and facilities of the three databases providing information on extragalactic objects via remote login. The status is that of mid-1994 and the underlined entries denote the database which in my opinion provides the best performance for a given application.

Given the sheer volume of the recent astronomical literature the retrieval of information on a given object via the classical printed volumes of abstract has become prohibitively time-consuming and thus explains the immense increase in the number of users of electronic databases. They have practically taken over the retrieval of bibliographical references in astronomy. It depends on the user's specific problem which database is the most appropriate to consult, and the user should be conscious of what level of completeness one can expect. If the user is interested in a list of "all" known objects in a small region of the sky, both SIMBAD and NED should be consulted. SIMBAD and NED are clearly more complete in bibliographic references to objects, but one should keep in mind that SIMBAD started only in 1983 to collect systematically the references for extragalactic objects, and NED's bibliography before 1988 is essentially that taken from SIMBAD. The bibliography of LEDA is independent of this and does not have this special limitation, but it offers only those references which actually contributed measured data to the database, whereas NED and SIMBAD attach to an object any reference mentioning the object, even if that object was only used as a calibrator or in a theoretical study. LEDA, in turn, has more data types to offer (up to 66 per object) and sample extractions can

Table 1: Comparison of the three major object-oriented databases

| Criterion | SIMBAD* | NED | LEDA |
|---|---|---|---|
| type of objects | all | <u>all</u> | galaxies, z<0.2 |
| number of objects | ∼80,000 | <u>∼300,000</u> | ∼100,000 |
| no. of data types | >20 | >20 <u>growing</u> | <u>≥60</u> |
| bibliography | good<br>since 1983 | <u>very good</u><br><u>espec. >1988</u> | only papers with<br>relevant data |
| search facilities and<br>extraction of samples | by all types<br>via "filters" | by object type<br>and redshift | <u>SQL on all</u><br><u>data types</u> |
| batch searches | yes | yes | yes |
| batch output format | complex | complex | <u>1 line/entry</u> |
| sky charts | (in prep.) | yes | <u>yes</u> |
| images | – | – | <u>POSS photos of</u><br><u>∼ 50,000 galaxies</u> |
| coord. systems, precession | <u>input+output</u> | <u>input+output</u> | in+output (SQL) |
| ease of use | manual recommended<br>(except X-interface) | easy | very easy |
| X-interface | yes | yes | yes |
| fee for use | yes (except US+F) | <u>no</u> | <u>no</u> |
| "Extras" | Dictionary Acronyms;<br>query by reference;<br>bibliographic tools | Acronyms;<br>Abstracts;<br>PhD abstracts;<br>Journal browser | Flamsteed<br>all-sky charts |

*) only the extragalactic content of SIMBAD is concerned here

be made by almost any combination of constraints on these parameters, but again, LEDA is limited to fairly nearby galaxies (up to z≈0.2). Each of the databases offers unique services not provided by the other two, e.g. SIMBAD allows a keyword search by subject through all the authors and titles of its bibliography, with NED it is possible to retrieve a list of all objects published in a given paper, while LEDA is currently the only one which offers images for many thousands of galaxies.

While these databases are of enormous help for extragalactic research, there are some aspects which remain to be worked on. One is the difficulty to find out *a priori* whether a certain publication has been completely folded into them or not. Also, there is a frequent misunderstanding that all entries in the CDS catalogues would automatically be accessible through SIMBAD (cf. sect. 4). Two of the more prominent examples of (electronically available) compilations folded *neither* into SIMBAD nor into NED, are the 9134 Zwicky clusters of galaxies [18], and the "Catalogue of Extragalactic Radio Source Identifications" [16]. The integration of the latter would be especially desirable since it gathers information on >10,000 sources drawn from ∼1000 references before 1983, the year when SIMBAD *started* to collect references on extragalactic objects (later taken over by NED). The lack of the Zwicky clusters occured to me after inspection of a Palomar Sky Survey print for the identification of an X-ray source. The source clearly coincided with a Zwickly cluster marked on the transparent POSS overlays as provided by Ohio State University. However, it took me some effort to convince a collaborator about the "reality" of this cluster as he could not "find" it neither in NED nor in SIMBAD . . . .
I noticed other examples of published papers in which objects were denoted as "unidentified" on the basis of NED or SIMBAD, while a consultation of [16] would have turned up a reference to a finding chart, published before 1983 !

SIMBAD and NED initially concentrated on gathering cross-identifications and bibliographic references, but they are incorporating more and more measured data (e.g. photometry) from the references. They are actually working towards a completion of literature coverage backwards in time. The above examples of lack of completeness are mainly intended to remind of other, complementary sources of information, and I discuss some of these below.

## 3. CATALOGUES AND COMPILATIONS

The largest electronic collection of astronomical catalogues is maintained at the CDS. A total of ∼900 data sets are archived, of which over 500 are available via anonymous FTP to the host `cdsarc.u-strasbg.fr`. However, only very few of these catalogues are completely integrated into SIMBAD, NED or LEDA. Therefore these catalogues represent a potential which is only little exploited at present [1]. Tools are being developed (e.g. SkyView [11], ESIS [8], ALADIN [13]) which allow to overlay objects from catalogues onto the digitised images from optical Schmidt surveys. The number of catalogues accessible like this is still very small and large efforts will be required to provide interfaces between the many different formats of catalogues and the search software.

An efficient updating of object-oriented databases requires electronic access to all newly (and relevant previous) published tabular data. I recently studied to what extent such tabular data are being preserved in electronic form [1]. My literature survey from 1987 to 1993 for articles with tabular information on at least 100 supposedly extragalactic objects revealed 374 such papers. As of May 1994 the CDS archive of electronic catalogues contained tables for 21 % of these, increasing from 11 % for 1987 to 29 % for 1993. Of the 60 papers with >1500 entries only half were in the archive. It appears that more of a collaborative effort between authors, journal editors and data centres is needed to collect such tabular material with a higher degree of completeness. Given that many electronic catalogues *not* in the CDS archive are available from various other services like ADS, DIRA2, EINLINE, ESIS, HEASARC, etc. (see [2]), a better coordination of the collection of these data sets and possibly a *Master Index of Electronic Astronomical Catalogues* are called for.

## 4. DATABASES FROM LARGE-SCALE SKY SURVEYS

At **radio** wavelengths two large-scale sky surveys are being performed with the Very Large Array (VLA). The *NRAO VLA Sky Survey* (NVSS) will map the whole sky north of $\delta = -42°$ at 20 cm with an angular resolution of $\sim 50''$ [5]. It will run from 1993 to 1996 as a service to the astronomical community, and the *uv* data and maps are publicly released immediately as they are reduced. The other survey, FIRST (*Faint Images of the Radio Sky at Twenty-cm*) will run from 1993 to 2000 to cover the north Galactic cap ($b > +30°$) at 20 cm and angular resolution of $\sim 5''$ [3]. In addition to the atlas of images both surveys will result in source catalogues with $\sim 2 \times 10^6$ entries down to their respective completeness limit of 2 and 1 mJy.

In the **far-IR** the IRAS all-sky survey will remain the reference for years to come. Apart from the various IRAS catalogues, extractions can be made from the atlas of images, e.g. via the *IRAS Postage Stamp Service* (URL `http://brando.ipac.caltech.edu:8888/ISSA-PS` on the World Wide Web).

The *Deep Near-Infrared Southern Sky Survey* (DENIS) [7] will cover the whole southern sky in three near-IR bands ($I, J, K$) using CCD technology. It will produce the first homogeneous catalogue of evolved galaxies complete up to z≈0.2 and is expected to detect altogether some 250,000 galaxies.

Ever increasing computer power as well as advances in the digitization of **optical** plates have made it possible to provide the all-sky atlas of Schmidt surveys as electronic images of a quality approaching that of the original plates. The best overview of efforts in this direction can be found in [10].

The most widely accessible of these digitizations is now the STScI scans of the red plates of the POSS I survey in the north, and the SERC J band survey in the south, available on 101 CD-ROMs distributed by the Astronomical Society of the Pacific. This project was the first of its kind and had been undertaken initially to construct the Guide Star Catalog for the HST. Public network access to this database is provided by SkyView (URL `http://skyview.gsfc.nasa.gov/skyview.html` on WWW).

Several other groups have undertaken similar digitizations of the Sky Surveys with better sampling, to produce object catalogues down to the plate limit, and object classification into stars or galaxies down to ∼1 mag above this limit. Both red (E) and blue (O) plates of POSS-I at high Galactic latitude ($|b| > 20°$) have been scanned by the APM ("Automated Plate Measuring") group at IoA Cambridge (UK) and also by the APS ("Automated Plate Scanner") at University of Minnesota. Both projects have produced catalogues of $\sim 10^9$ objects, of which a few percent are classified as galaxies. Small portions of the catalogues can be extracted and finding charts produced from them by remote access. While the APM catalogue goes to fainter magnitudes (with probably more spurious objects at the faint end), the APS also offers the images of detected objects. Numerous discrepancies in the classification of individual objects go along with the different approaches.

The southern counterpart of these has been prepared using the COSMOS plate scanning machine at the Royal Observatory Edinburgh. The scans were made from the IIIa-J and Short Red Surveys, and cover the sky south of $\delta = +2.5°$. Access to the object catalogue of several $10^8$ objects will be provided through the Anglo-Australian Observatory (AAO) from late 1994 [2]. A new digitization of the southern sky survey using the SuperCOSMOS machine will give an even better quality object database, which is planned to be made available on CD-ROM by 1995.

Scanning of the second epoch Palomar (POSS-II) survey has begun at STScI, and the object catalogue is being prepared at Caltech [17].

The Sloan Digital Sky Survey (SDSS) [9] will produce a digital photometric map of $1\pi$ sr around the north Galactic pole down to $23^m$. The object catalogue is expected to contain $10^8$ galaxies and $10^6$ quasar candidates. Of these, $10^6$ galaxies and $10^5$ quasars will be selected to obtain high resolution spectra.

At shorter than optical wavelengths where photons become rarer and more expensive to detect, our knowledge of the extragalactic source population is less profound. Source lists from **extreme UV** surveys contain only very few extragalactic objects (e.g. [4]). In the soft **X-ray** regime the ROSAT satellite surveyed the whole sky but the source catalogue with ∼60,000 sources has not yet become publicly available. From the many hundreds of pointed ROSAT observations a further catalogue of maybe ∼200,000 serendipitous and much fainter sources may be extracted over patches of the sky.

# 5. CONCLUSION AND OUTLOOK

While there is much duplication of effort in maintaining various object-oriented databases, one must admit that not only were different databases designed for different purposes, but also that a minimum redundancy will help to improve their quality. The preparation of databases, compilations and reference catalogues is a painstaking and longterm task. Although very much needed for fundamental studies, it receives insufficient financial support from funding agencies. If future compilations and reference catalogues are to be drawn more efficiently from electronic databases, we must expend a greater effort on quality control of these databases.

Clearly, one challenge of the future will be the cross-identification of objects detected in the various wavebands. Many cross-identifications between catalogues in various wavebands are already being made *silently* by SIMBAD, NED and LEDA during their regular updates (see e.g. [15]). It would be both technically and scientifically undesirable if all objects from large-scale surveys were entered into one single object-oriented database. For the millions of faint objects drawn from plate scans we are too far still from a reliable automatic classification, while the brighter and extended galaxies are often 'taken into pieces' by the object extraction software. Automated cross-identification algorithms naturally introduce some spurious identifications. Efficient links should thus be created between object-oriented databases, astronomical catalogues and the bulky large-scale surveys to allow the users to check the available information on individual objects in all possible detail.

Very sophisticated tools for graphical overlays of object catalogues and digitised optical sky surveys are now becoming available, which will allow both optical and cross-identification of objects detected in various wavebands. However, we should spend a comparable effort in the preservation of published material in electronic form. While we have been rather unconcerned with this in the past, the future of the publishing business will bring in all likelihood an electronic access to both data and text. The available *tools* will easily display the information for us on the screen, but only if we have this information readily *available*.

**Acknowledgements** – I am grateful to G. Paturel for suggesting this contribution and for the hospitality I received over the past year in the LEDA team at the Observatoire de Lyon.